# Bifunctional Metamaterials with Simultaneous and Independent Manipulation of Thermal and Electric Fields

*Chuwen Lan[1,2#], Yuping Yang[3#], Xiaojian Fu[4#], Bo Li[2]\*, Ji Zhou[1]\**

Dr. C. Lan, Prof. J. Zhou
State Key Laboratory of New Ceramics and Fine Processing, School of Materials Science and Engineering, Tsinghua University, Beijing 100084, China
E-mail: zhouji@mail.tsinghua.edu.cn
Prof. L. Bo, Dr. C. Lan
Advanced Materials Institute, Shenzhen Graduate School, Tsinghua University, Shenzhen, 518000, China
E-mail: boli@mail.tsinghua.edu.cn
Prof. Y.P. Yang
School of Science, Minzu University of China, Beijing, 100081, China.
Dr. X. Fu
State Key Lab of Millimeter Waves, School of Information Science and Engineering, Southeast University, Nanjing 210096, China

Metamaterials offer a powerful way to manipulate a variety of physical fields ranging from wave fields (electromagnetic field, acoustic field, elastic wave, etc.), static fields (static magnetic field, static electric field) to diffusive fields (thermal field, diffusive mass). However, the relevant reports and studies are usually conducted on a single physical field or functionality. In this study, we proposed and experimentally demonstrated a bifunctional metamaterial which can manipulate thermal and electric fields simultaneously and independently. Specifically, a composite with independently controllable thermal and electric conductivity was introduced, on the basis of which a bifunctional device capable of shielding thermal flux and concentrating electric current simultaneously was designed, fabricated and characterized. This work provides an encouraging example of metamaterials transcending their natural limitations, which offers a promising future in building a broad platform for manipulation of multi-physics field.

Usually, the realization of physical functionality means to manipulate and control the relevant physical fields in a desired way. To accomplish such a goal, various structures and geometries are designed based on available materials. However, the capacity and functional diversity are usually limited by the material it is made of. As a result, some desirable properties seem impossible to achieve with materials available at present. For example, a material with independently controllable thermal and electric conductivity is difficult to come by. Once such material becomes available, numerous intriguing possibilities will be opened up and myriad novel applications will find their way into our life.

The rapid development of metamaterials is beginning to bring such a bright future ever closer to us. Metamaterials actually are composites consisting of well-arranged subwavelength inclusions that can be considered as "effective" material. Compared with conventional materials which are synthesized at the molecular level, the novel conception of metamaterials are capable of tailoring material properties at a subwavelength level. What is more interesting, metamaterials derive their properties not from materials used in the process of fabrication, but from their designed structures. With all these novelties, metamaterials possess numerous exotic properties and promise many applications. Over the past years, metamaterials were widely used in manipulation of physical fields ranging from electromagnetic wave,[1-3] acoustic wave,[4] elastic wave[5] to matter wave.[6] Recently, metamaterial was introduced to control fields such as static magnetic field,[7-11] *dc* electric field,[12-14] thermal field,[15,16] electrostatic field[17] and diffusive field.[19-20] However, bifunctional or multifunctional metamaterials are rarely reported and studied. It is interesting to explore the possibility of creating novel metamaterials to achieve independently controllable thermal and electric conductivity. Recently, Moccia introduces "transformation multiphysics" framework and concept of bifunctional metamaterial, and a shell consisting of thousands of thermal and electric elements was theoretically designed to act as thermal concentrator and electrical invisibility cloak.[21] Such study provides an encouraging example of metamaterials transcending their natural limitations, which may open up novel possibilities in the largely unexplored phase space of multifunctional/ multi-physical devices, and realize considerable potential applications. Up to date, however, the corresponding experimental demonstration is still unexplored. This might be attributed to the fact that the transformation-based devices require anisotropic, gradient and extreme parameters, which greatly complicates the fabrication. Consequently, it is highly desirable to explore the possibility of a simpler scheme and give the corresponding experimental demonstration.

Here, we proposed and experimentally demonstrated a bifunctional metamaterial to manipulate thermal and electric fields simultaneously and independently. A composite with simple structure was designed with independently controllable thermal and electric conductivity. Based on this composite, we designed, fabricated and characterized a simple bifunctional device capable of shielding thermal flux and concentrating electric current simultaneously, which confirms the feasibility of our scheme.

The key to manipulating the thermal and electric fields independently is constructing a medium with independently controllable thermal and electric conductivity. To accomplish such goal, we introduced a bifunctional metamaterial composed of fan-like inclusions in the associated ($\rho$, $\phi$, $z$) cylindrical coordinate system, as shown in Figure 1a. A fan-like unit cell with periodicity of $p\rho \times p\phi$ is illustrated in the magnified details, which consists of several fan-like inclusions made of four types of materials (A, B, C, D). The thermal and electric conductivity for materials A, B, C, D are ($\kappa_A$, $\sigma_A$), ($\kappa_B$, $\sigma_B$), ($\kappa_C$, $\sigma_C$) and ($\kappa_D$, $\sigma_D$) respectively. Here, electroinsulating material-- Acrylonitrile Butadiene Styrene (ABS) with $\sigma_A$=0 S/M is used as background material (material A). Due to its poor thermal conductivity ($\kappa_A$=0.25W/mK), it is treated as a thermal insulation ($\kappa_A$=0W/mK) for the sake of simplicity. An aluminum (material D, $\kappa_D$=207W/mK, $\sigma_D$=2.1e7 S/M) fan-like inclusion with a size of $a \times \alpha$ is placed in the center. To manipulate thermal and electric fields independently, one can employ two special materials (material B and material C). Material B is aluminum nitride (AlN), which has a high thermal conductivity ($\kappa_B$=190W/mK) and good electrical insulation performance ($\sigma_B$=0 S/M). Other materials with these properties can also be used here, for instance, ceramics (beryllium oxide, boron nitride, aluminum oxide), intrinsic semiconductor materials (silicon, germanium). As is shown in this picture, material-B-based fan-like inclusion (with a size of $\alpha$ in $\phi$ direction) is placed along $\rho$ direction and touches the material D. Material-C inclusion is made of conductive silver adhesive, which has good electric conductivity ($\sigma_C$=5.5e5 S/M) but poor thermal conductivity ($\kappa_C$=1W/mK). Such inclusion (with length of $a$ in $\rho$ direction) is placed along $\phi$ direction and touches the material-D inclusion. Clearly, this composite with such cell would show /display good thermal conductivity in the $\rho$ direction and poor thermal conductivity in $\phi$ direction. Meanwhile, it has good electric conductivity in $\rho$ direction and good electrical insulation performance in $\phi$ direction. To calculate the effective thermal and electric conductivity tensors, we consider the conductive silver adhesive as thermal insulation ($\kappa_2$=0W/mK) for simplicity. According to effective medium theory (EMT),[22] one can get the thermal and electric conductivity tensors:

$$\kappa_\rho = \frac{p\rho \kappa_B \kappa_D}{(p\rho - a)\kappa_D + a\kappa_B} \cdot \frac{\alpha}{p\phi} \quad (1)$$

$$\kappa_\phi = m \quad (2)$$

$$\sigma_\rho = 0 \quad (3)$$

$$\sigma_\phi = \frac{p\phi \sigma_C \sigma_D}{(p\phi - \alpha)\sigma_D + \alpha \sigma_C} \cdot \frac{a}{p\rho} \quad (4)$$

Here, $m$ is much smaller than $\kappa_\rho$ ($m \ll \kappa_\rho$). Apparently, $\kappa_\rho$ is dependent on $p\rho$, $p\phi$, $\kappa_A$, $\kappa_D$, $a$, and $\alpha$, but independent of $\kappa_C$. Meanwhile, $\sigma_\phi$ is dependent on $p\rho$, $p\phi$, $\sigma_C$, $\sigma_D$, $a$, $\alpha$, but independent of $\sigma_B$. Additionally, $\sigma_\rho = 0$ and $\kappa_\phi = m$, therefore, one can design the effective thermal and electric conductivity tensors independently by adjusting $\kappa_B$, $\sigma_C$, $a$, $\alpha$, $p\rho$, $p\phi$. Consequently, by using such composite, one can manipulate the thermal and electric fields independently.

As a typical example, a bifunctional device capable of shielding thermal flux and concentrating current simultaneously is designed in this work. Figure 1b-e shows the schematic diagram of the proposed bifunctional device. Suppose that the space is divided into three parts: interior region ($r<R1$), shell ($R1<r<R2$), exterior region ($r>R2$) (Figure 1b). The thermal and electric conductivity for background medium (interior and external regions) are $\kappa_0$ and $\sigma_0$, while those for the shell are $\kappa_1$ and $\sigma_1$, respectively. In Figure 1c, the thermal flux flows from left to right, and the electric current is produced from left to right. As shown in Figure 1d, the thermal flux flows circumvent the interior region, keeping their original paths without any distortions. The distortion of thermal fluxes only occurs in the shell, indicating that a thermal cloaking effect is achieved. As for electric domain (see Figure 1e), the current touching the shell is concentrated into the interior region, resulting in increased current density, while the exterior current keeps its original path without distortion. In other words, the shell functions as electric concentrator. To implement such a device, one can employ the so-called "transformation multiphysics".[21] In such a scheme, transformations are applied to thermal and electric fields simultaneously, and the required parameters can be achieved with a bifunctional metamaterial. However, such a scheme suffers from gradient and extreme parameters. In addition, the requirement of independently controllable thermal and electric conductivity makes it even more difficult. Therefore, here we employ anisotropic but homogeneous medium. Assume that the background material is homogeneous but with anisotropic thermal conductivities $\kappa_r$, $\kappa_\theta$. Here, make $\kappa_r \cdot \kappa_\theta = \kappa_0^2$, to ensure that the external thermal flux is kept undistorted. Then, the only remaining task is to make thermal flux flow around the interior region. We introduce a variable $c$, where $c=\kappa_r/\kappa_\theta$. According to the work,[23] one can find that when $c<9/100$, a nearly perfect cloak can be obtained. Moreover, the smaller $c$ corresponds to the better performance. Similarly, as for electric current, we assume that the shell is made of homogeneous but anisotropic electric material with $\sigma_r$, $\sigma_\theta$, where $\sigma_r \cdot \sigma_\theta = \sigma_0^2$. According to the previous work,[11] when $\sigma_r > \sigma_\theta$, the current density in the interior region is higher than the exterior one. The smaller value of $c$ ($c=\kappa_\theta/\kappa_r$) leads to better concentrator efficiency. Based on the above analysis, using anisotropic but homogeneous metamaterial will greatly simplify the corresponding realization of the function.

Based on the aforementioned theories, we have designed such a bifunctional device using bifunctional metamaterial illustrated in Figure 2a. Here, silicon carbide (α-SiC) with a thermal conductivity of $\kappa_0$=20W/mK and an electric conductivity of $\sigma_0$=1.0 S/M is chosen as the background material. The outer and inner radii of the shell are set as 17mm and 5mm, respectively. The shell is composed of arrayed cells made of several specific materials: silver paste ($\kappa_2$=1 W/mK, $\sigma_2$=5.5e5 S/M), ABS ($\kappa_0$=0.25 W/mK, $\sigma_0$=0 S/M), aluminum ($\kappa_3$=207 W/mK, $\sigma_3$=2.1e7 S/M), aluminum nitride ($\kappa_1$=190W/mK, $\sigma_1$=0 S/M). According to the effective medium theory (EMT)[22] and simulation, the corresponding optimized geometry parameters are: $p\rho = 4mm$, $p\phi = 20°$, $a = 3mm$, and $\alpha = 10°$. The fabricated sample is shown in Figure 2b. First, simulations are carried out to obtain the temperature profile distribution. For comparison, the case for homogeneous background material is also simulated and shown in Figure 3a. As expected, a uniform temperature gradient is generated from left to right. The corresponding thermal flux distribution is provided in Figure 3c, where a uniform value can be obtained. The simulation results for our bifunctional device are shown in Figure 3b,d. From Figure 3b, it can be found that the exterior temperature profile distribution has not been altered with the presence of bifunctional device, and the distortion for the isothermal lines only occurs inside the bifunctional device. As seen in Figure 3d, the thermal flux in the inner region is reduced by a factor of 0.03, while that in the exterior region remains unchanged. The results given above show that this device can function as a thermal cloak. To demonstrate its capacity of concentrating the electric current in the inner region, 1V potential is applied between the two sides, as schematically illustrated in Figure 1c. For comparison, we first simulate the potential distribution and current density distribution in homogeneous background material, as shown in Figure 3e,g. As expected, the isopotential lines are straight and paralleled to each other, and the current densities are uniform. The simulation results of potential distribution and current density for our bifunctional device are provided in Figure 3f and Figure 3h. First, the exterior isopotential lines remain straight and parallel to each other with no distortions (Figure 3f). Second, the isopotential lines touching the device are compressed in the inner region. This leads to the increased current densities in the inner region, which can be verified by the simulated current density distribution as shown in Figure 3f, where the current density in the inner region is enhanced by a factor of 2. Consequently, this device can function as an electric current concentrator.

In the experiment, the left side of sample is put into contact with hot water (333 K), and the right side is with ice water (273 K). To characterize the thermal property, an infrared heat camera (Fluke Ti300) is employed to measure the temperature profile distribution. To reduce the reflection of sample for the operating wavelengths of the thermal heat camera, thin electrical insulation tape with emissivity higher than 95% is attached to the surface of the sample. The measured results are presented in Figure 4a-b. Clearly, the temperature gradient in the core region is decreased greatly and the external field is nearly undistorted. Good agreement can be obtained between the simulation and measured results. To experimentally confirm the electric property, a Multimeter (Agilent 34410A, 6, 1/2Digit Multimeter) is used to obtain the potential distribution. The performance of electric concentrator can be evaluated by the potential distribution along the lines $x$=-20mm, $x$=20mm and $y$=0, as given in the insets of Figure 5. Clearly, the isopotential lines (both for line $x$=-20mm and $x$=20mm) in homogeneous background material are straight and those for our device also remain straight due to the fact that there is no distortion in the external field. As for $y$=0mm, the potential gradient is considerably enhanced in our device compared to that in homogeneous background material, indicating a good electric current concetration effect for our device. The simulation and experimental results show good agreement, confirming the feasibility of our design.

At this point, it has been demonstrated that by employing bifunctional metamaterial, the thermal and electric fields can be manipulated simultaneously and independently. As for the designed bifunctional device, anisotropic but homogeneous thermal and electric parameters were designed to concentrate or cloak physical field. Compared with TO method, our method can avoid many problems, such as gradient and extreme parameters, thus greatly simplifying the fabrication process. It is noteworthy that the proposed bifunctional metamaterial can also be used for "transformation multiphysics". This is because that our scheme is a general method for designing desired independently controllable thermal and electric parameters. It's also worth mentioning that our method can be applied in other bifunctional devices.

In conclusion, we have provided the theoretical and experimental realization of simultaneous and independent manipulation of thermal and electric fields with bifunctional metamaterials. We introduced a bifunctional metamaterial with independently controllable thermal and electric conductivity to achieve independent manipulation of thermal and electric fields. Based on this concept, we demonstrated a bifunctional device capable of shielding thermal flux and concentrating electric current simultaneously. This work provides a novel approach towards independently tailoring material properties, therefore promising a broad platform for manipulation of multi-physics field.

**Supporting Information**
Supporting Information is available from the Wiley Online Library or from the author.

**Acknowledgements**

This work was supported by the National Natural Science Foundation of China under Grant Nos. 51032003, 11274198, 11574408, 51221291 and 61275176, National High Technology Research and Development Program of China under Grant No. 2012AA030403, Beijing Municipal Natural Science Program under Grant No. Z141100004214001, and the Science and technology plan of Shenzhen city undergrant Nos. JCYJ20120619152711509 , JC201105180802A and CXZZ20130322164541915. Author 1, Author 2 and Author 2 contributed equally to this work.
[1] R. A. Shelby, D. R. Smith, S. Schultz, *science* **2001**, *292*, 77.
[2] D. Schurig, J. Mock, B. Justice, S. A. Cummer, J. B. Pendry, A. Starr, D. Smith, *Science* **2006**, *314*, 977.
[3] N. Landy, S. Sajuyigbe, J. Mock, D. Smith, W. Padilla, *Phys. Rev. Lett.* **2008**, *100*, 207402.
[4] S. Zhang, C. Xia, N. Fang, *Phys. Rev. Lett.* **2011**, *106*, 024301.
[5] M. Farhat, S. Guenneau, S. Enoch, *Phys. Rev. Lett.* **2009**, *103*, 024301.
[6] S. Zhang, D. A. Genov, C. Sun, X. Zhang, *Phys. Rev. Lett.* **2008**, *100*, 123002.
[7] B. Wood, J. Pendry, *J. Phys.: Condens. Matter* **2007**, *19*, 076208.
[8] F. Magnus, B. Wood, J. Moore, K. Morrison, G. Perkins, J. Fyson, M. Wiltshire, D. Caplin, L. Cohen, J. Pendry, *Nat. materials* **2008**, *7*, 295.
[9] S. Narayana, Y. Sato, *Adv. Mater.* **2012**, *24*, 71.
[10] A. Sanchez, C. Navau, J. Prat-Camps, D. X. Chen, *New J. Phys.* **2011**, 13, 093034.
[11] C. Navau, J. Prat-Camps, A. Sanchez, *Phys. Rev. Lett.* **2012**, *109*, 263903.
[12] F. Yang, Z. L. Mei, T. Y. Jin, T. J. Cui, *Phys. Rev. Lett.* **2012**, *109*, 053902.
[13] W. X. Jiang, C. Y. Luo, H. F. Ma, Z. L. Mei, T. J. Cui, *Sci. Rep.* **2012**, 2.
[14] F. Yang, Z. L. Mei, X. Y. Yang, T. Y. Jin, T. J. Cui, *Adv. Funct. Mat.* **2013**, *23*, 4306.
[15] S. Guenneau, C. Amra, D. Veynante, *Opt. Express* **2012**, *20*, 8207.
[16] S. Narayana, Y. Sato, *Phys. Rev. Lett.* **2012**, *108*, 214303.
[17] R. Schittny, M. Kadic, S. Guenneau, M. Wegener, *Phys. Rev. Lett.* **2013**, *110*, 195901.
[18] C. Lan, Y. Yang, J. Zhou, B. Li, *arXiv preprint arXiv:* **2014**,*1412*, 3294.
[19] S. Guenneau & T. M. Puvirajesinghe, *J. R. Soc. Interface* **2013**, *10*, 20130106.
[20] L. Zeng, R. Song, *Sci. Rep.* **2013**, *3*, 3359.
[21] M. Moccia, G. Castaldi, S. Savo, Y. Sato, V. Galdi, *Phys. Rev. X* **2014**, *4*, 021025.
[22] D. J. Bergman, *Phys. Rep.* **1978**, *43*, 377.
[23] T. Han, T. Yuan, B. Li, C. Qiu, *Sci. Rep.,* **2013,** *3*, 1593**.**

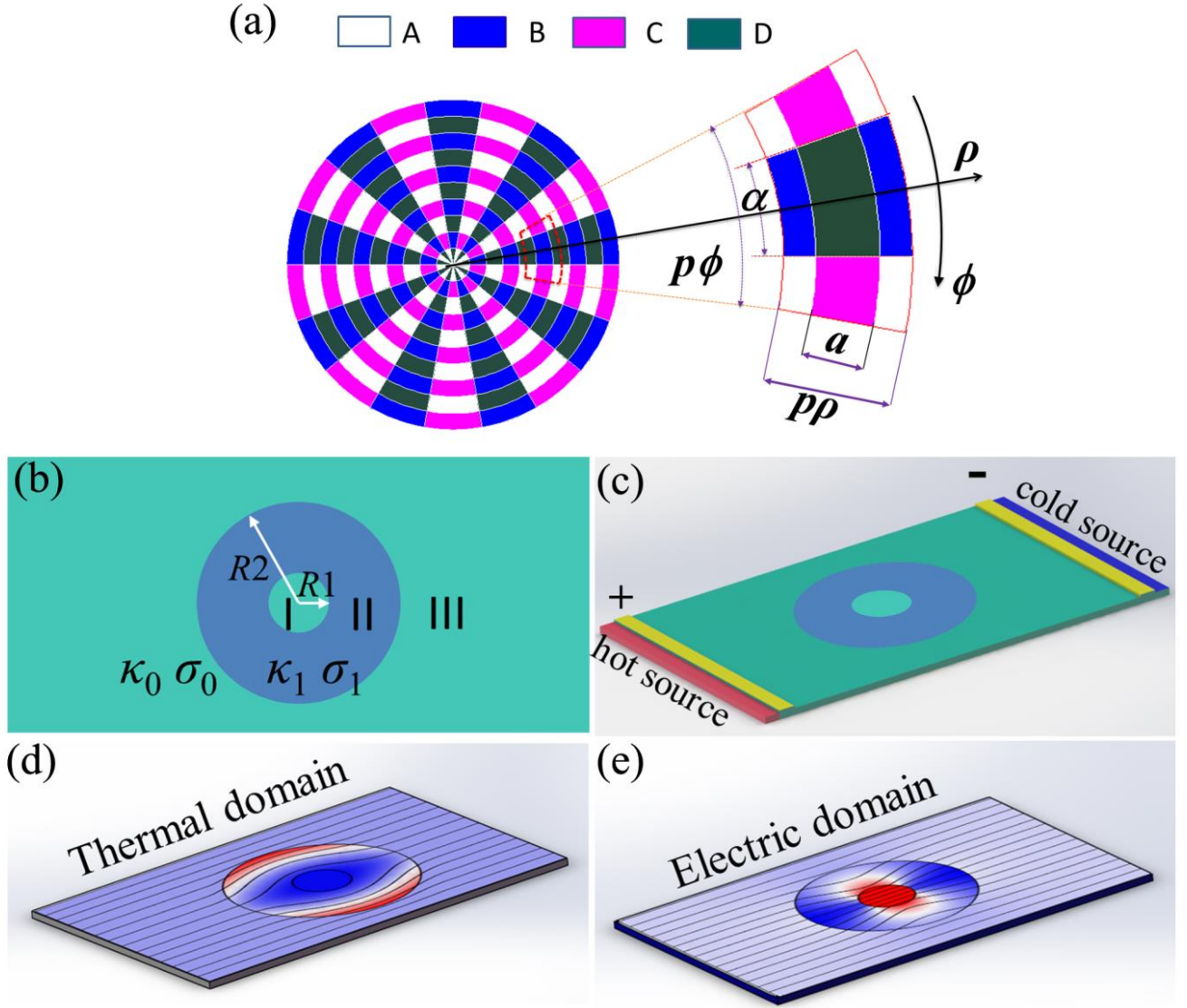

Figure 1. (a) The proposed bifunctional metamaterial with independently controllable thermal and electric conductivity in the associated ($\rho$, $\phi$, $z$) cylindrical coordinate system. The unit cell is composed of several fanlike inclusions made of material A, B, C, D. The thermal and electric conductivity for materials A, B, C, D are ($\kappa_A$, $\sigma_A$), ($\kappa_B$, $\sigma_B$), ($\kappa_C$, $\sigma_C$) and ($\kappa_D$, $\sigma_D$) respectively. The corresponding geometrical parameters can be seen in the insert. The principle for bifunctional device behaving as thermal cloak and electric concentrator: (b) The corresponding physical model. the space is divided into three parts: interior region ($r<R1$), shell ($R1<r<R2$), exterior region ($r>R2$) (see Figure 1b). The thermal and electric conductivity for background medium (interior and external regions) is $\kappa_0$, $\sigma_0$, while the one for the bifunctional shell is $\kappa_1$, $\sigma_1$, respectively. (c) The bifunctional device applied with temperature

gradient and electric potential gradient. (d) The thermal flux distribution. (e) The current distribution.

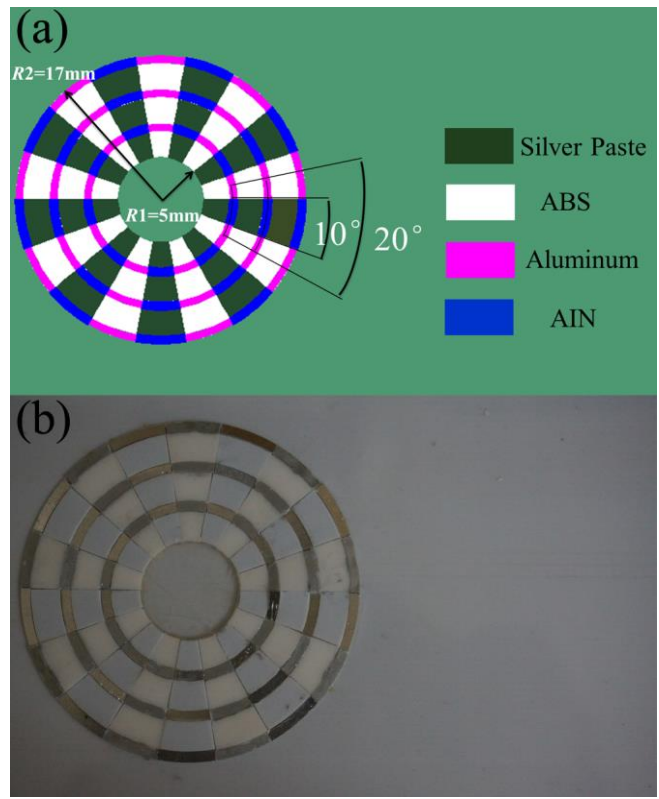

Figure 2. (a) The schematic illustration for practical realization of bifunctional device. the corresponding geometry parameters are optimized as follow: $p\rho$=4mm , $p\phi = 20°$, $a = 3mm$, $\alpha = 10°$. (b) The photogragh for the fabricated sample.

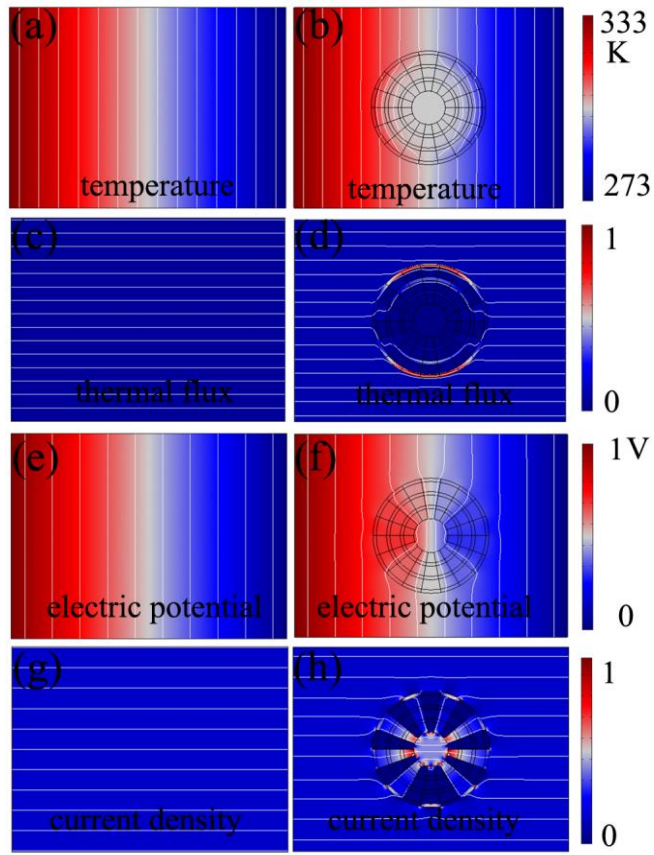

Figure 3. Thermal simulation results for background material: a) temperature profile. c) thermal flux distribution. Thermal simulation results for bifunctional device: b) temperature profile. d) thermal flux distribution. Electric simulation results for background material: e) electric potential distribution. g) current density. Electric simulation results for bifunctional device: f) electric potential distribution. h) current density.

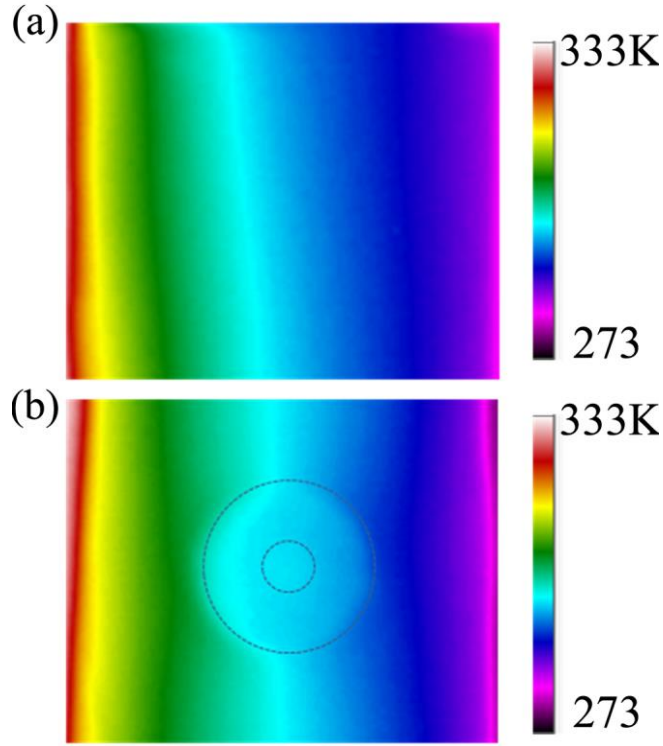

Figure 4. a) Measured temperature profile for background material. b) Measured temperature profile for bifunctional device.

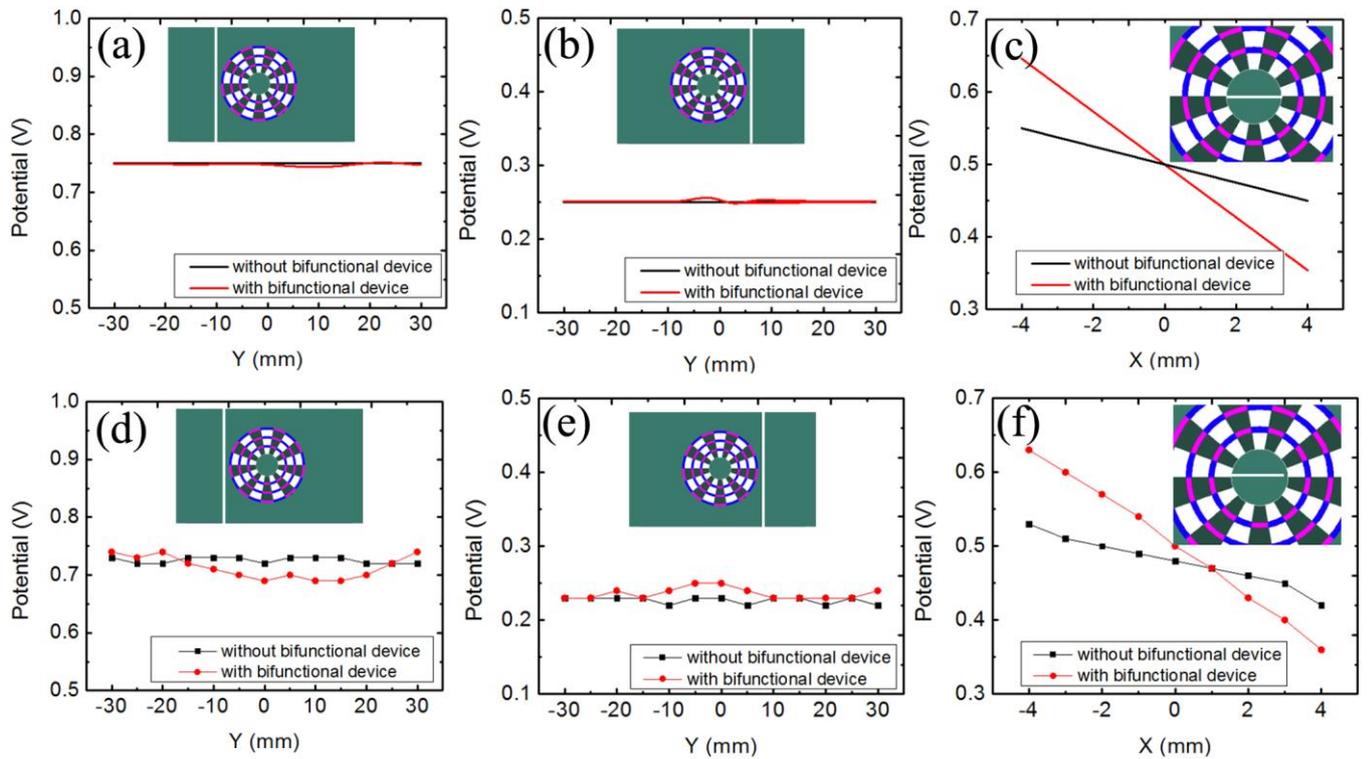

Figure 5. The simulation and experiment results for background material and the one with electric concentrator: simulated electric potential values for the different cases at corresponding positions: (a) $x$=-20mm, (b) $x$=20mm and (c) $y$=0mm. d, e, f) Corresponding experimental potential values, respectively. The white lines in inserts represent observed lines.